# Paper 01

# Using Dynamic Simulation in the Development of Construction Machinery

Reno Filla [1] and Jan-Ove Palmberg [2]

[1] Volvo Wheel Loaders AB, Eskilstuna, Sweden
[2] Dept. of Mechanical Engineering, Linköping University, Sweden

## Abstract

As in the car industry for quite some time, dynamic simulation of complete vehicles is being practiced more and more in the development of off-road machinery. However, specific questions arise due not only to company structure and size, but especially to the type of product. Tightly coupled, non-linear subsystems of different domains make prediction and optimisation of the complete system's dynamic behaviour a challenge. Furthermore, the demand for versatile machines leads to sometimes contradictory target requirements and can turn the design process into a hunt for the least painful compromise. This can be avoided by profound system knowledge, assisted by simulation-driven product development. This paper gives an overview of joint research into this issue by Volvo Wheel Loaders and Linköping University on that matter, and lists the results of a related literature review. Rather than giving detailed answers, the problem space for ongoing and future research is examined and possible solutions are sketched.

*A witty quote proves nothing.*

*(Voltaire)*

# 1  Introduction

The general motives for "Virtual Prototyping" are probably familiar to all engineers: Stricter legal requirements (e.g. with regard to exhaust emissions and sound) and tougher customer demands (e.g. with regard to performance and handling) lead to more advanced, complex systems, which are harder to optimise. With traditional methods, development will cost more and need more time. In contrast to this, increased competition demands lower development cost and shorter project times.

"Virtual Prototyping" has been generally adopted in the vehicle industry as a major step towards solving this conflict both on the consumer side (cars) and on the commercial side (trucks and buses). Having started with simulation of sub-systems, the state-of-the-art is simulation of complete vehicles, mostly for evaluation of handling, comfort, and durability but also for crash-tests.

One reason for the off-road equipment industry lagging behind can be found in the size of these companies: being significantly smaller, broad investments in the latest CAE tools (together with the necessary training) take longer until amortisation. The other, and probably more important reason is that the products are very different to those of the on-road vehicle industry – not only geometrically (size), but topologically (sub-systems of various domains and their interconnections).

Cases have recently been published where complete machines were simulated for evaluation of the simulation technique itself [1], sub-systems [2], comfort-related aspects [3], or durability [4]. This paper too, will deal with dynamic simulation of complete machines, but for analysis and optimisation of overall performance and related aspects. The focus will be on wheel loaders with hydrodynamic transmissions, but most findings (and questions) will be also applicable to other off-road machinery.

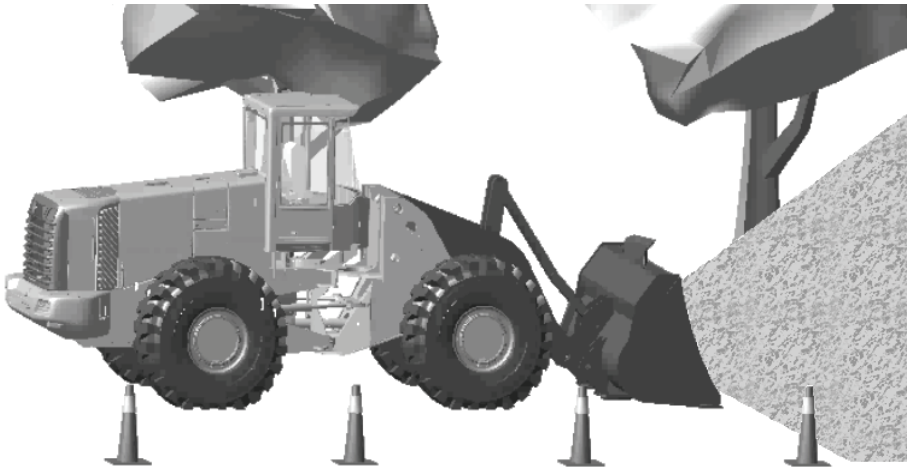

**Figure 1.** Multi-body model of a Volvo L220E wheel loader, loading gravel

## 2  Background

Due to the versatility of these machines, wheel loaders need to fulfil a great many requirements, which are often interconnected and sometimes contradict each other. This is true of essentially all industrial products and is widely recognised. Since this paper focuses on performance (and related issues), aspects such as total cost of ownership, market availability, reliability etc. will not be discussed explicitly.

To give just a few examples, the following performance-related aspects are important (varying with the working task):

- Geometric parameters (lift height, digging depth, dump reach, parallel alignment)
- Loads, torques and forces (tipping load, break-out torque, lift force, traction force)
- Speeds and cycle times (complete machine and sub-systems)
- Consumption and emissions (fuel consumption, exhaust emissions, sound & vibration)
- Controllability (precision, feedback, response)

While some of the items are clearly determined by more than just one sub-system (e.g. lift force, which is determined by hydraulics and load unit), others seem to be possible to attribute to one single sub-system (e.g. traction force). One might thus, wrongly, be tempted to leave such aspects out of the optimisation loop when it comes to trading-off product targets against each other when choosing technical solutions.

A modern wheel loader of hydro-dynamic design, however, consists of tightly coupled, non-linear sub-systems of different domains. Since all sub-systems interact even in seemingly simple cases, prediction and optimisation of the complete system's dynamic behaviour is a challenge.

Figure 2 shows how the sub-systems of main interest are interacting when loading gravel.

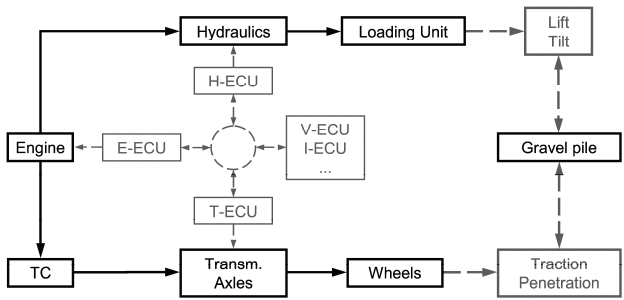

**Figure 2.** Simplified transfer scheme of a wheel loader, loading gravel
(TC = Torque Converter, ECU = Electronic Control Unit, V = Vehicle, I = Instrument)

For loading granular material like gravel, the bucket first has to penetrate the pile. This requires traction force, which is achieved by transferring torque from the engine via a torque converter, transmission, axles, and wheels to the ground. A typical sequence for actually filling the bucket is then to break material by tilting backwards a little, lifting a little, and penetrating even further. The lift and tilt functions require engine torque to be transferred via hydraulic pumps (converting torque to hydraulic pressure), cylinders (converting hydraulic pressure to longitudinal force), and loading unit.

As shown, the two different transfer paths compete for the limited engine torque. Furthermore, in the act of loading, these two paths are brought together in the gravel pile: the penetration establishes a reaction force at the bucket, which counteracts the break-out force, as well as the lift force. Like a short circuit, loads can be transferred back to the origin, in this case the engine, and lead to overload (greater interaction between the systems or the engine stops completely).

This has to be avoided by a design that carefully balances traction and lift/tilt. But these functions can not be optimised without influencing others, e.g. machine speed and lifting time. In addition to this, Figure 2 neglects the fact that loading unit geometry and bucket design are of great importance for a smooth bucket filling. In addition, a machine operator who handles a loader in an unintelligent way can have difficulty in achieving a favourable balance. In such cases (but not exclusively), implementing sophisticated electronic control strategies can be of great help.

Above, one single phase in one single (yet frequent) handling case has been described. A typical loading cycle consists of several phases, where balance has to be established in each. Additionally, there are a great many different handling cases, each with its own requirements that need to be satisfied.

With the ambition of developing a product that is significantly better than its predecessor every time, this becomes harder and harder to achieve. The trend towards more electronic control is as striking in the off-road equipment industry as it is in the on-road branch (passenger cars, trucks, and busses). In Figure 2, this is symbolised by five electronic control units, each serving a specific purpose (component or sub-system). The ring represents the common data bus (CAN-bus) the ECUs are connected to.

## 3   Design Process, Current Position

Since this research project is being carried out at Volvo Wheel Loaders, parts of their product development process will be used as examples throughout this paper. However, there are several indications that the procedure is similar for other companies in the off-road industry.

The intention is not to describe the entire development process, but rather to focus on the parts that handle overall machine performance, as described in the previous section.

After setting up preliminary product targets, an initial calculation loop is started, with the objective of finding a good balance between loading unit, drive train (including en-gine) and hydraulics. In this process, product targets are gradually refined, until the overall system's specifications are judged acceptable and feasible (Figure 3).

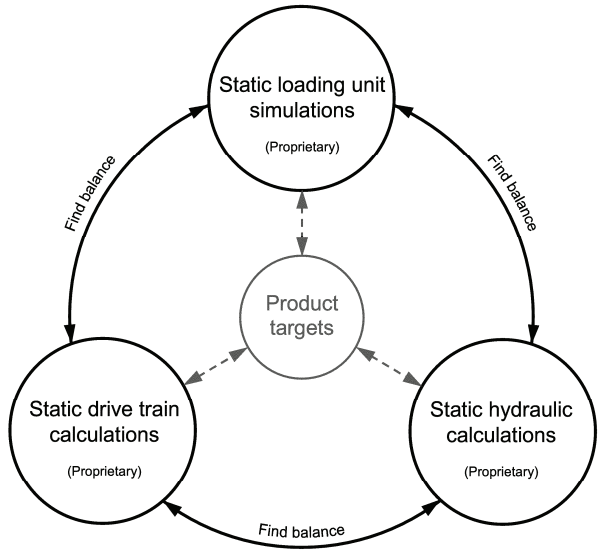

**Figure 3.** Initial calculation loop for balancing the main systems

After this step, which involved several assumptions, derivative analysis is performed for all major sub-systems, which will either confirm those assumptions or require a return to the first loop with modified input parameters.

As sketched in section 2, a wheel loader (a complex system of tightly coupled, non-linear sub-systems) shows complex (non-linear) behaviour in complex (real-life) situations. To optimise the first one, the latter two need to be understood as fully as possible. In this case, there are multiple handling situations, each involving humans with complex behaviour of their own. Since the initial loop only relies on static calculations (though guided by valuable experience), there is the possibility that the supposedly found optimum will not work flawlessly in a dynamic reality.

It is therefore praxis to build early, so-called "functional prototypes" as the next step. Using them for extensive real-life tests, the intention is to uncover any static and dynamic problem at an early stage in the design phase, where reiteration is still inexpensive. However, these functional prototypes are often not built with exactly the same components as planned for the production machines, since those are still in development and therefore only available as early prototypes (or not at all). This poses the risk that dynamic shortcomings will not be detected until later.

One example is that engines have been found to be critical. Electronically controlled engines can have higher response times than older, mechanically controlled ones. There are several reasons for this, one being that the ECU itself needs some time to collect all the information from the sensors, process it, and calculate the appropriate action. Additionally, modern engines have to be certified according to the latest exhaust emission legislation. Virtually no diesel engine today shows any sign of much visible smoke in transient load situations. This is achieved by limiting the injected fuel quantity at low turbo pressures, thus making the engine a little less responsive. A second example is hydraulic pumps: modern wheel loaders consist of load-sensing hydraulic systems, which in most cases are superior to the old open-centre design. In such LS-systems the response time of the pump is a critical factor. If a functional prototype is built with one or more pumps with incorrect dynamic behaviour, together with a diesel engine as described above, the complete loader's dynamics can differ significantly from those of a series-manufactured one.

A requirement for the next stage, building "real" prototypes, is that all major components are at least mature prototypes, close to series production. In this stage, all major testing is done and detailed design enters its final phase. Due to the above mentioned procedure for functional prototypes additional dynamic complications are sometimes detected. This is unfortunate, but can be handled by evaluation and preparation of possible adjustments in advance.

In the next stage, building a so-called "production prototype", all components have to be series manufactured. Originally, this prototype was intended to allow the production facility to test and refine assembly methods. But since even mature prototypes of major components might show a different dynamic behaviour to those from series production, this last prototype can also be used for one more validation. Any detected, performance-related issue in this stage is very costly and can jeopardise the complete project time schedule – but building pre-series machines with shortcomings in the next step is unacceptable. Testing of this prototype (and its now updated predecessors) is therefore very thorough, i.e. resource-consuming.

In general, evaluation of machine performance consists of several stages, such as component validation (e.g. engine), sub-system testing (e.g. lift time and traction force), and finally testing of the complete loader. The latter is first performed by experienced test personnel at the company's proving ground. Having passed those tests, the prototypes are sent to selected customers, who use them in their everyday work. This ensures that a greater variety of operators have the chance to judge the design.

Such a procedure is often referred to as a "V-approach", where global targets are broken-down into targets for sub-systems and components (top-down), with the validation being done the other way around (bottom-up). It is interesting that the more the testing resembles real life, the less measurable the outcome becomes. Specifications of isolated components are easily validated by measurements in test benches. Sub-systems are a little harder to test, mainly because there are few test benches that are appropriate, but also because not every aspect of the dynamic behaviour has been specified. Testing complete machines by professional test personnel sometimes reveals that these people are far more experienced than the average wheel loader operator. The risk of being biased, even involuntarily, exists. Despite this, with a limited proving ground and limited personnel, it is impossible to take all variations of handling case and operating habit into account. Therefore, the machines are field-tested by selected customers. However, in the last two scenarios the validation of complete machine performance relies strongly on the operator and his/her subjective judgement. While the employed test personnel are more used to expressing their impression in measurable terms, it is the voice of the customer that counts in the market. The challenge is how to quantify the criteria for a well-balanced machine. And how to quantify all boundary conditions during a measurement, i.e. how to ensure repeatability.

## 4  Design Process, Vision

The aim of the present project is *evolution* of the current product development process, rather than *revolution* by means of Design Science (see [5] for a critical review by Frost). The research question is therefore how to augment the existing design process with dynamic simulation. As mentioned before, the focus is on analysis and optimisation of overall performance and related aspects.

The revised design process has to fulfil the following non-optional targets: Lead to development of

… products of at least equal high performance, efficiency, and operability

… but with increased robustness

… in a shorter time

… at a lower development cost

compared with today. Saved resources (time, money, and people) can then be spent on optimising one or all of the aspects mentioned in the first item in the list.

In an earlier project, a valuable lesson has been learned: speed matters when it comes to iterations, especially in the concept phase. When Volvo's old loading unit calculation programme was to be replaced by a more modern version, this was done with a proprietary simulation system, which was based on a multi-body system (MBS) and a modern database. The development was done in-house. This new simulation system has proved to be more flexible, more accurate, and especially more efficient for the user, except for some pre-study engineers, who used the superior speed of the old calculation programme to brute-force optimise loading unit geometries. Since the new system obtains results by multi-body simulation, rather than calculation of hard-coded explicit equations, one run takes a couple of seconds longer than with the old programme. Brute-force optimisations of the old type are no longer time efficient. If this had been known before, i.e. if it had been included in the project targets, one could have developed a special downscaled version, that was less accurate but faster. Introduction of the new system forced those pre-study engineers to abandon a time-efficient technique that worked well.

A similar risk can be seen with this research project: the current initial (static) calculation loop (Figure 3) is fast and reasonably accurate. The shortcoming today is rather that the dynamic behaviour of the complete machine is first evaluated by testing a func-tional prototype, followed by testing a "real" prototype. Therefore a moderately revised process (as shown in Figure 4) is proposed.

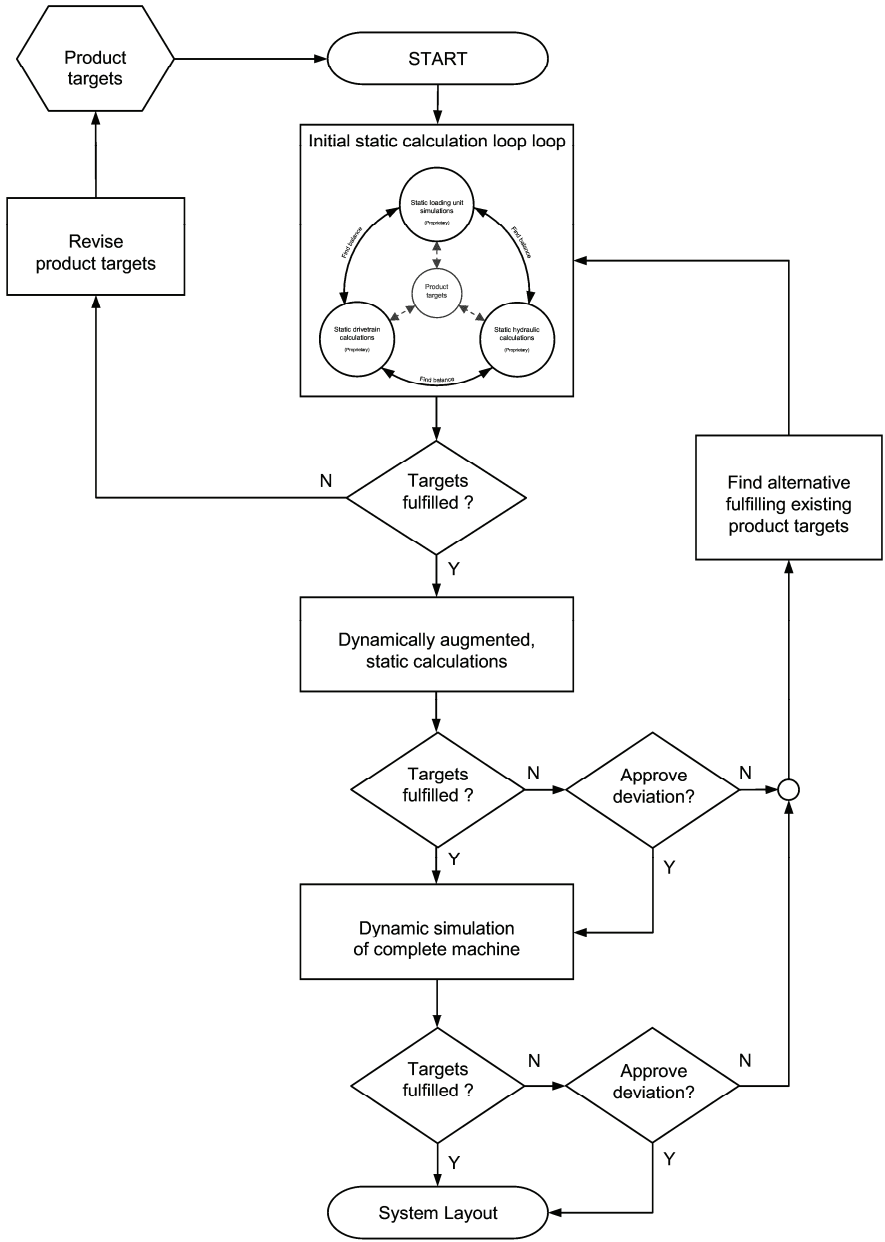

**Figure 4.** Revised design process

As practised today, the process starts with feeding the product targets into the initial static calculation loop. If no satisfactory solution can be found, the targets need to be revised. The next step is described as "Dynamically augmented, static calculations" (see below for an explanation). Here, as well as in the next step "Dynamic simulation of complete machine", non-fulfilled product targets do not necessarily lead to a reiteration, as long as the deviation can be approved. Since product targets will probably never cover all product properties (including dynamic behaviour), this checkpoint will give the whole process some flexibility. Only if the deviation is too high will a new static calculation loop need to be started (with the original product targets as input). If no alternative layout candidate can be calculated, the product targets need to be revised.

Since simulating complete machines can take a long time, a quick check on whether the solution from the initial loop is dynamically feasible, would be of great help. This is supposed to be done in the second step "Dynamically augmented, static calculations". Concrete methods still need to be developed. However, one example can be given (applies to machines with hydrodynamic transmissions and LS-hydraulics): a critical phase in a so-called "short loader cycle" (or V-cycle) is when the machine, coming backwards with a full bucket, changes direction towards the load receiver (e.g. an articulated hauler or a dump truck). During that time, there is a close interaction between the main subsystems:

1. In order to reverse the machine, the operator lowers the engine speed (otherwise the gear shifting will be jerky and the transmission couplings might wear out prematurely). Less torque is available at lower speeds. Additionally, the engine response is worse at lower speeds, mainly due to inertia of the turbo charger (and smoke limiter settings, as explained earlier).

2. When switching gears from reverse to forward, the loader is still rolling backwards. This forces an abrupt change of rotational direction of the torque converter's turbine wheel, thus greatly increasing the slip, which leads to a sudden increase in torque demand from the engine.

3. Some operators do not stop lifting the loading unit while reversing, thus requiring high oil flow throughout the whole process. The oil flow is proportional to the hydraulic pump's displacement and shaft speed, and the load-sensing pumps can be assumed to be directly connected to the engine crankshaft. Thus, at lower engine speeds and high demand of oil flow, the displacement goes towards maximum (simply put). The amount of torque that a pump demands of the engine is proportional to its displacement and hydraulic pressure. Since the loader's bucket is full and due to the loading unit's geometry, hydraulic pressure is high. With displacement at maximum, this too leads to an increased torque demand.

Both drive train and hydraulics thus suddenly apply higher load to an already weakened engine. All depends on the time scale of these three concurrent phenomena, which is why a satisfactory answer can only be given by a detailed dynamic simulation (or testing of a real machine, i.e. a functional prototype). However, an approximate, less time-consuming first approach is possible: given the loader speed when switching gears from reverse to forward, and given the engine speed at that time, the maximum slip between the pump wheel and the turbine wheel of the torque converter can be calculated, and thus the maximum demanded engine torque (by using the torque converter's specifications). In the worst case scenario, hydraulic pressure and pump displacement can be assumed to be maximal. Together with the engine speed, this gives the second engine torque demand. If the sum of both torque demands is larger than the available steady-state engine torque at that speed, the proposed system layout will almost certainly lead to dynamic problems. If the available steady-state engine torque is considerably larger than the sum of both torque demands, the system will most probably function as intended. To check the case in between, it is important to consider that due to factors such as turbo charger inertia and smoke limiter mentioned before, an accelerating engine seldom has full steady-state torque at lower engine speeds. Therefore, checking against the *static* torque curve might give a false sense of security.

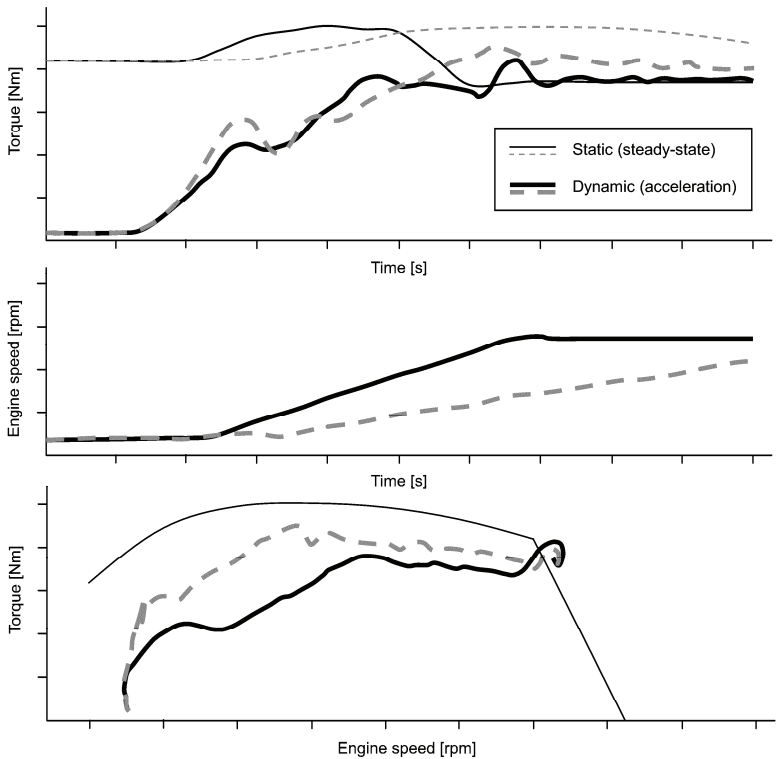

**Figure 5.** Dynamic torque curve of a modern, turbo-charged diesel engine. Less gap between static and dynamic torque at lower acceleration (grey, dashed curve) due to faster boost pressure build-up relative to engine speed

What is needed is a *dynamic* engine torque curve, measured at a typical acceleration rate. Assuming the engine is released to low idle during reversing (worst case); the acceleration rate is simply obtainable as the speed difference from the engines max torque speed to low idle, divided by a typical time for the reversal phase. Using a test bench with an electro-magnetic brake, an engine run-up with a forced acceleration as described above can be performed. The amount of available torque during that phase equals the necessary braking torque (Figure 5).

In Figure 5, the grey, dashed curves represent a test, where the engine was allowed to accelerate less. In theory, this should not only generate more available torque because the rotating parts consume less torque for acceleration, but also because there is more time for turbo boost pressure to be built up. As can be seen in the first diagram, the engine torque itself does not increase faster *in time* for the case with less forced acceleration. Instead, confirming the theory, more torque is available *relative to engine speed*, expressed as a smaller gap between static and dynamic torque curve for the slower acceleration. If the engine were allowed to accelerate sufficiently slowly, the dynamic torque curve (bold line style) would follow the static one (thin line style).

In order to quickly sort out system layout candidates that will lead to dynamic problems, more such checking methods need to be developed, together with refinement of the existing rules of thumb.

Taking a critical look at the proposed revised design process, even if it has been labelled as "evolutionary rather than revolutionary" in the beginning of this section, it implies a bigger change than might be obvious at first glance. With its introduction, functional prototypes are only built to verify system layouts that have passed both steps that check the dynamic behaviour. These steps are given the great responsibility of deciding whether a certain system layout can be further pursued or is to be sent back for reiteration. The worst scenario would be to reject cases that actually would work in reality, but nobody will ever know since only approved cases enter the functional prototype phase. To avoid this and to build confidence, there has to be a smooth transition between current practice and the proposed process.

In the literature, this is referred to as the change from simulation-verified design to simulation-driven design. Campbell shows the usual charts and gives some practical tips on this in [6]. T Larsson examines this topic from a research perspective in [7]. Even though Murphy and Perera are writing about their findings on Discrete Event Simulation, their paper [8] provides some valuable insight, such as classifying "Team building" as one major contribute factor for success, while "Software selection" is only mentioned as a minor factor. They also list the activity of making effective internal promotion for simulation as a major success factor, advice that is similar to the findings of [9], where Beskow and Ritzén present the findings of their research on change management in general.

# 5 Puzzle Pieces

In order to fulfil the vision drafted in the previous section, progress needs to be made in various areas. Some pieces of this puzzle have already fallen into place, though.

## 5.1 Simulation and Modelling Techniques

As shown earlier, off-road equipment consists of systems from various domains, and most of them need to be taken into account when simulating performance of complete machines – which is usually a collaborative activity. As noted by many researchers, engineers have often already chosen one domain-specific simulation program that they are familiar with. Instead of forcing migration to one monolithic simulation system that can be used in several domains (but offers only limited functionality in the individual domain), a better approach is to couple the specialised, single-domain tools. This has the advantage that both pre- and post-processing are done decentralised, in the engineer's domain-specific tools. In [10] J Larsson develops a technique for such co-simulation and applies it to a model of a complete wheel loader. The present research project will use this approach and further develop it.

This paper focuses more on the early stage of product development, but simulation is equally beneficial in all stages. As T Larsson notes in [7], the more knowledge is gained, the more the level of detail increases on the way from concept development over detail design to product support. Being able to reuse most of the model in all stages demands a modular approach. With that, individual modules can be replaced with updated versions (e.g. for increasing the level of detail for one specific sub-system) without having to make changes to the surrounding modules. Once the model of the complete machine has been divided into modules, communication between them is then only allowed through their public interfaces. Thus, an update of a module's internal structure does not affect the surrounding modules, since the interface is not changed. In multi-body systems, such an interface usually consists of some markers and transfer variables. But in some special cases, other descriptions of module interaction such as surface contact or field interaction need to be modelled (if the simulation tool does not provide a feature for this). Sinha et al. write in [11] about the need for modelling interaction dynamics between modules as modules of their own. In the present research project though, this should be avoidable by carefully selecting the module interfaces so that they are still physically motivated and intuitive, yet allow rigid coupling of markers and variables.

A hierarchical modular approach combined with co-simulation will give the engineer great flexibility regarding model creation. An engine module for instance could be modelled as a true multi-body module with pistons, piston rods, crankshaft, rotational damper, flywheel etc. (combustion pressure to be modelled as a variable force acting on the piston). This would give the possibility to study torsional vibrations throughout the drive train. Or the engine could just be approximated as a set of explicit MBS-equations acting on a flywheel (i.e. no additional 3D-parts are modelled). Combining this with co-simulation, the MBS-equations could be replaced by a Matlab/Simulink model of the engine, which delivers the torque value to the multi-body system, where it is applied to the flywheel. In either case, other drive train modules will always connect to the flywheel and thus not be structurally affected by switching between different engine modules. However, their behaviour and that of the complete system are both affected by the switch, since the different engine modules in this example have different levels of detail.

Looking at the relation of *level of detail* to modelling, Finn and Cunningham observe in [12] that "the process of constructing a mathematical model is an inexact activity, which is not supported by any underlying theory or governing laws". Others go so far as to equate mathematical modelling with art, which is quite an uncomfortable thought for an engineer. Brooks and Tobias published a comprehensive overview in [13] and provide valuable guiding considerations on *level of detail* and *complexity*: "The purpose of a complexity measure is to characterise the model so that this information can aid the choice of model by predicting model performance. Ideally we would like to have a single, system independent definition and measure of complexity covering all the aspects of the level of detail of a model and applicable to all conceptual models. However, no such definition or measure exists and as a result the term complexity itself is a source of confusion due to its usage in many different contexts." Such as using this term for anything that someone finds difficult to understand. The authors further note that "a single absolute measure of model performance cannot be obtained, but a meaningful comparison of alternative models in similar circumstances should be possible". However a search of the available literature reveals that little progress has apparently been made since these observations (1996).

Another aspect of modular modelling is that flexibility regarding input data procurement is gained. Planning to reuse an existing model by parameter update implies that exactly the same type of input data is available for the update, as it was for the original model. If a hydraulic pump has been modelled in great detail with hundreds of parameters, then an update is difficult if all the engineer has got is a characteristics curve. "The more detailed a model is, the more time must be spent on defining the parameters, based on measurements or theoretical engineering reasoning", notes Makkonen in [14]. Test data might not be compatible due to different measurement set-ups, and theoretical reasoning requires the engineer to reason in a similar (if not identical) way to the creator of the original model. In practice, it might prove more time and cost efficient to just develop a new module from scratch, but with an identical interface.

Later on, assembly and replacement of modules could be done semi-automatically using macros. However, it is dangerous to create too rigid procedures that let the engineer out of the loop. In order to replace just one module in a system, the new module has to have a similar level of detail, otherwise the entire system's behaviour is affected. Which level of detail will be chosen for the various modules is dependent on the phenomena to be studied; an experienced engineer's knowledge is vital in this decision process and should not be overruled by some automatic routine.

### 5.2 Models

Naturally, since the goal is simulation of complete machinery for the purpose of analysis and optimisation of overall performance (and related aspects), most of the machines sub-systems and components need to be modelled in a modular way and in various levels of detail. Despite some issues with input data procurement (see above), this is relatively easy to achieve. The challenge lies in those parts of the system that are beyond the company's control: operator and environment.

The research on car driver models is significant and involves topics such as lane change manoeuvres and path tracking. Recently, the research on driver models for articulated heavy vehicles has increased. However, wheel loaders are a very different kind of machinery and we are unlikely to find a suitable off-the-shelf operator model. Loader operators need to maintain the delicate balance between hydraulics and drive train. Interaction with other road-users is non-existent (the only exception being the need to match a certain time frame in order to co-operate effectively with operators of other construction equipment, such as haulers or excavators). The driver model does not need to include proper reaction to unforeseen events (parking car, red light etc.); and visual orientation can be modelled roughly, because it can be assumed that a professional driver knows the workplace very well. Nechyba presents his research in [15], demonstrating how to model human control strategies from real-time human control data. His application of these for a continuous model uses a flexible cascade neural-network combined with extended Kalman filtering. For the discontinuous case, control actions are modelled by individual statistical models. Kiencke et al. introduce a hybrid driver model in [16], which describes the human perception process by discrete event techniques and can be adapted to different driver types. The driver model, made in SES/Workbench, was then evaluated in a co-simulation with a Matlab/Simulink vehicle model.

The development of in-depth operator models is in itself a substantial task and will be left out of the discussion in this paper. However, using the company's extensive knowledge of wheel loader operation in general, it is not too difficult to extract some basic rules, which can, for example, be represented either by a model in Matlab/Simulink or by means of IF/STEP-statements and timers directly in an MBS model. Some promising results have already been obtained in this way.

Another important part of the overall system is a proper environment model. For most work scenarios, this might be trivial to achieve. However, this is not the case for bucket loading of gravel or other granular material. In [17] Singh presents the state-of-the-art in automation of earth moving. Related to this are the studies made by Hemami in [18] on the motion trajectory during bucket loading of an LHD-loader (Load-Haul-Dump). A review of resistive force models for earthmoving processes is made by Blouin et al. in [19]. In [20] Ericsson and Slättengren present a model for predicting digging forces. This model, used at Volvo Wheel Loaders, is realised as a general force subroutine in the multi-body system ADAMS and has been verified with measurements of cylinder pressures from excavation of coarse gravel.

### 5.3 Simulation Goals

Simulation-driven design implies using *analysis* to provide insight for *synthesis*. This requires one to specify the goal of the conducted simulation. Besides the already mentioned expectations to reduce the cost and duration of product development, important goals for Volvo Wheel Loader with this project are

- Optimisation of machine performance
- Optimisation of efficiency (with fuel efficiency as one aspect)
- Optimisation of operability (which driveability is just a subset of)
- Optimisation of a design's robustness

These goals are at the same time overlapping and partly contradictory. In order to solve this conflict, techniques for multi-objective optimisation will be needed – a research area that Andersson gives an account of in [21]. This also includes a study on achieving robust design by means of a genetic algorithm. A wealth of papers exists on tolerance analysis and design of experiment techniques. Simpson et al. present in [22] a review of such meta-modelling techniques as RSM (response surface methodology) and Taguchi; but at the same time they warn that many applications using these methods are "statistically questionable, because many analysis codes are *deterministic* in which error of approximation is *not* due to random effects".

Optimisation applied to a simulation of a complete machine like a wheel loader could either be optimisation of the technical system or optimisation of the driver model, i.e. to answer the question of how a machine of a given design should be operated to achieve the highest fuel efficiency or highest performance. Optimisation of robustness is here not limited to geometric tolerances, but includes variances of all possible parameters, even such as workplace altitude (the barometric pressure heavily affects the response of engines) or experience of the machine operator.

The term *operability* includes *driveability* as a subset, valid for the drive train only. Operability, a term that later on in this project needs to be defined, considers the interplay of all sub-systems of the complete machine and includes such aspects as how well-balanced an operator *perceives* a certain machine to be, or how efficiently a work cycle can be performed. As touched upon in section 3, the operator's subjective judgement needs to be quantified in order to find an appropriate objective function for optimisation. Wicke presents his research on driveability and control aspects of vehicles with Continuously Variable Transmissions in [23]. Insight gained there should be applicable at least to the driveability-part of off-road machinery's operability. Quantifying the latter also includes developing real-world test procedures that are compatible with those in the virtual world, and vice versa.

# 6  Conclusion and Outlook

This paper presented the joint research by Volvo Wheel Loaders and Linköping University on simulation of complete machines for analysis and optimisation of overall performance. The motivation on the side of the industrial partner is to develop products of equally high performance, efficiency, and operability, but with more robustness regarding these aspects, in a shorter time and at a lower total development cost. A revised product development process (with regard to the research topic) has been proposed. Examples of areas for future research have also been presented.

Research in the immediate future will focus on a definition of operability (including quantification), as well as practical simulation problems (such as defining modules and physically motivated and intuitive interfaces between them).

In a longer perspective, a measure of complexity needs to be developed. Research will be done on how to optimise operability, which will also include a look at control strategies.

18  *Paper 01*